\documentclass[conference,a4paper]{IEEEtran}

\usepackage{graphicx}
\usepackage[caption=false,font=footnotesize]{subfig}
\usepackage{amsmath,amssymb}
\usepackage{siunitx}
\usepackage{balance}
\usepackage{url}

\begin{document}

\title{Single-Feed Circularly Polarized \\Super Realized Gain Antenna}

\author{
\IEEEauthorblockN{Georgia Psychogiou\IEEEauthorrefmark{1},
Donal P. Lynch\IEEEauthorrefmark{2},
Spyridon N. Daskalakis\IEEEauthorrefmark{3},\\
Manos M. Tentzeris\IEEEauthorrefmark{4}
George Goussetis\IEEEauthorrefmark{3},
and
Stylianos D. Asimonis\IEEEauthorrefmark{1}}
\IEEEauthorblockA{\IEEEauthorrefmark{1}Department of Electrical and Computer Engineering, University of Patras, Greece}
\IEEEauthorblockA{\IEEEauthorrefmark{2}Centre for Wireless Innovation (CWI), Queen's University Belfast, United Kingdom\\}
\IEEEauthorblockA{\IEEEauthorrefmark{3}
	 School of Engineering and Physical Sciences, Heriot-Watt University, EH14 4AS Edinburgh,, United Kingdom
}
\IEEEauthorblockA{\IEEEauthorrefmark{4}
	School of Electrical and Computer Engineering, Georgia Institute of Technology, Atlanta, GA 30332 USA
}
Email: dlynch27@qub.ac.uk, s.daskalakis@hw.ac.uk, s.asimonis@upatras.gr}

\maketitle

\begin{abstract}
This paper presents a super realized gain, circularly polarized strip-crossed dipole antenna operating at 3.5~GHz. Superdirective behavior is achieved by leveraging strong inter-element mutual coupling through careful adjustment of the strip dimensions. The antenna features a single driven element, with the other element passively loaded with a reactive impedance. The structure is optimized to maximize left-hand circularly polarized (LHCP) realized gain, ensuring high polarization purity and good impedance matching. The optimized design exhibits a 50~$\Omega$ impedance bandwidth of 3.29--4.17~GHz (23.75\%) and an axial-ratio bandwidth of 3.43--3.57~GHz (4\%). At 3.5~GHz, the antenna achieves a peak realized gain of 6.1~dB ($ka \approx 1.65$), with an axial ratio of 1.4~dB. These results demonstrate that circular polarization and superdirectivity can be simultaneously realized in a geometrically simple, low-profile ($0.15\lambda$) antenna, rendering it suitable for integration into compact sub-6~GHz wireless and sensing platforms.
\end{abstract}


\section{Introduction}

Superdirectivity is a classical yet continually relevant phenomenon in antenna theory, describing the ability of an array of closely spaced radiating elements to achieve directivity values substantially higher than those expected from its physical aperture \cite{Hansen1981Fundamental}. In an idealized form, such performance arises from precise amplitude and phase control among the array elements, resulting in highly constructive interference in the end-fire direction and near-complete cancellation elsewhere \cite{uzkov}. However, practical realization of superdirectivity has long been constrained by issues including extreme sensitivity to phase and amplitude errors, high reactive energy storage, and narrow impedance bandwidth \cite{Assimonis2023How}. These challenges have historically limited superdirective designs to theoretical studies and experimental prototypes rather than practical implementations.

Recent advances in full-wave simulation, numerical optimization, and high-precision fabrication have re-established superdirectivity as a viable design goal for electrically small antennas (ESAs) \cite{Altshuler2005monopole, Yaghjian2008esefa}. Modern optimization algorithms, particularly multi-objective and evolutionary methods, now enable efficient exploration of the trade-space between directivity, efficiency, and bandwidth, allowing for the synthesis of compact arrays that approach theoretical directivity limits without excessive complexity \cite{jiao2018antenna, Goudos2013moa}. The renewed interest in superdirective arrays is also driven by the growing demand for compact, high-performance antennas in wireless systems operating at sub-6 GHz frequencies, where physical aperture constraints are particularly severe \cite{ibrahim2023design}.

In parallel, circular polarization (CP) has become increasingly important across satellite communication, global navigation, and emerging 5G/6G applications \cite{chen20235g}. CP provides enhanced link stability by reducing sensitivity to polarization mismatch and mitigating multipath fading in dynamic propagation environments. Among the wide range of CP antenna configurations, crossed-dipole structures remain attractive for their mechanical simplicity, polarization purity, and ease of fabrication \cite{ta2015crossed, yepes2019analysis}. When two orthogonal dipoles are excited with equal amplitudes and a 90$^\circ$ phase difference, either left-hand (LHCP) or right-hand circular polarization (RHCP) can be achieved with minimal complexity \cite{toh2003}.

The present work combines these two concepts, superdirectivity and circular polarization, within a single compact end-fire array architecture. The proposed single-feed antenna comprises two closely spaced crossed-dipole elements optimized to operate at 3.5 GHz, building upon the related works \cite{Lynch2024,lynch2024compactmeandereddipolearray}. Through strong mutual coupling (inter-element distance of $0.15\lambda$), careful adjustment of the dipole dimensions, and the introduction of a reactive load on the non-driven element, the current phase distribution is manipulated to reinforce radiation in the end-fire direction. This enables a compact, circularly polarized, and superdirective response without the need for external phase-shifting or matching networks.

The antenna was analyzed and optimized using ANSYS HFSS \cite{ansys_hfss}, employing its built-in multi-objective genetic algorithm (MOGA) to maximize LHCP realized gain, aiming for both impedance matching and circular polarization purity. The resulting design demonstrates that compact arrays ($\approx ka$=1.65, where $ka$ is the electrical size of the antenna) can approach theoretical directivity bounds while maintaining practical bandwidth and efficiency. Moreover, the proposed structure highlights the feasibility of achieving circular polarization and superdirectivity simultaneously in a geometrically simple low-profile configuration, suitable for integration within space-constrained wireless platforms such as Unmanned Aerial Vehicle (UAV), Internet of Things (IoT), Multiple Input Multiple Output (MIMO) front-ends.

\section{Antenna Configuration and Design}
\label{sec:config}

\begin{figure}[!t]
\centering
\includegraphics[width=0.99\columnwidth]{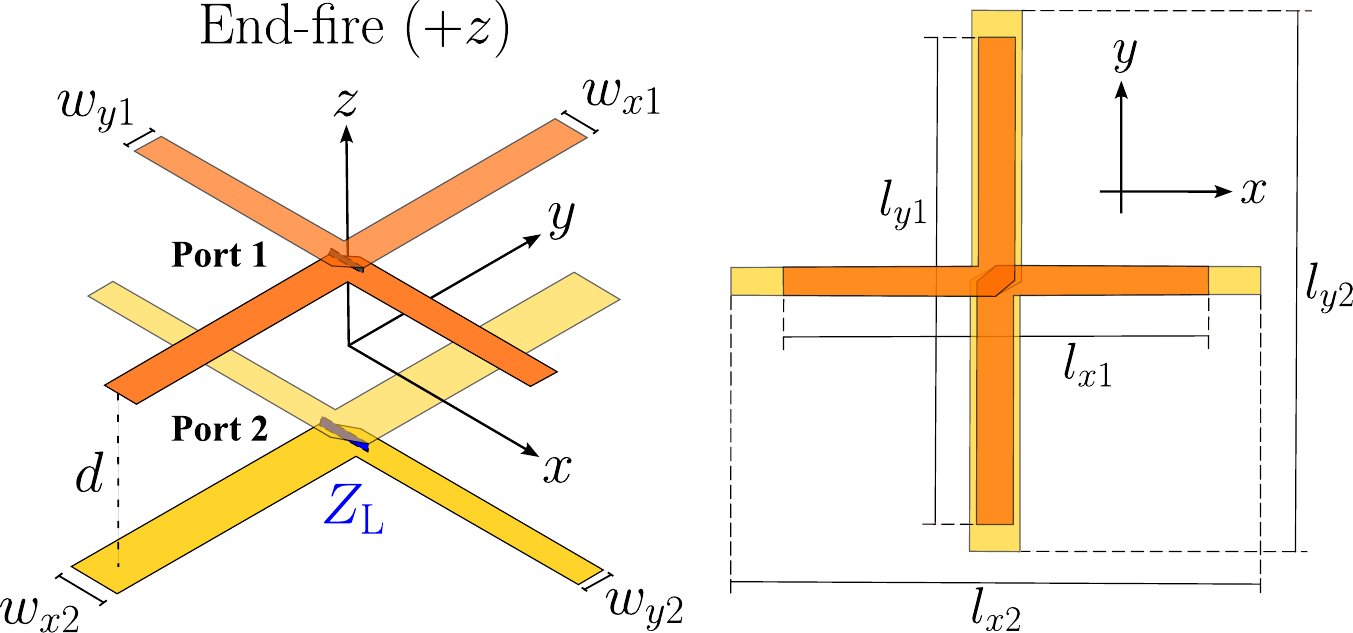}
\caption{Geometry of the optimized two-element crossed-dipole array showing all design parameters ($l_{x1}$--$l_{x2}$, $l_{y1}$--$l_{y2}$, $w_{x1}$--$w_{y2}$) and inter-element spacing $d$ along the $z$-axis. Each element is differentially fed at its center, and the lower element includes the reactive load $Z_L$.}
\label{fig:geometry}
\end{figure}

The proposed antenna employs a two-element crossed-strip dipole configuration to realize circularly polarized end-fire radiation with superdirectivity. Each element consists of two orthogonal dipoles stacked along the $x$- and $y$-axes and excited with equal amplitudes and a $\pm90^\circ$ phase difference to generate LHCP or RHCP. The dipoles are modeled as thin metallic strips suspended in free space.
As illustrated in Fig. \ref{fig:geometry}, the configuration comprises two crossed-dipole elements arranged in parallel planes along the $z$-axis, separated by a center-to-center distance of $d = 0.15\lambda$. This spacing was identified during optimization as providing a favorable balance between impedance matching to 50~$\Omega$, super realized gain, and circular polarization, while maintaining a compact end-fire length.

In this antenna design, only one crossed dipole (the upper element) is differentially excited (feeding point), while the lower element acts as a parasitic radiator. In a realistic implementation, a balun would likely be employed to ensure proper balanced feeding from an unbalanced transmission line. The orthogonal arm geometry produces the required in-quadrature fields for LHCP without external hybrid networks. The LHCP configuration can be easily switched to RHCP by incorporating a phase shift component to reverse the quadrature phase offset. To achieve end-fire radiation, the lower parasitic element is reactively loaded with an impedance $Z_L$. The reactive component introduces a controlled phase shift between the coupled currents of the two radiators, reinforcing radiation along the forward ($+z$) direction and suppressing the backward lobe.

The optimization of the proposed antenna was performed in ANSYS HFSS using the MOGA framework. The goal of the optimization process was to maximize the LHCP realized gain in the end-fire $+z$ direction at the target frequency of 3.5\,GHz. This metric inherently combines the effects of impedance matching, radiation efficiency, and circular polarization purity, since the realized gain accounts for both mismatch and ohmic losses. Therefore, the optimization problem can be formulated as follows:

\[
\begin{aligned}
\text{Maximize:} \quad & 
G_{\text{LHCP}}\big(l_{x1}, l_{x2}, l_{y1}, l_{y2}, w_{x1}, w_{x2}, w_{y1}, w_{y2}, Z_L\big) \\
\text{Subject to:} \quad &
\begin{cases}
l_{i} \in [0.3,\,0.6]\,\lambda_0, \\
w_{i} \in [0.005,\,0.05]\,\lambda_0, \\
Z_L \in [-5000,\, +5000]~\Omega,
\end{cases}
\end{aligned}
\]
where $G_{\text{LHCP}}$ represents the realized gain of the left-hand circularly polarized component obtained directly from the far-field results of the simulation. The parameters $l_{x1,2}$ and $l_{y1,2}$ denote the arm lengths of the crossed dipoles along the $x$ and $y$ axes, respectively, while $w_{x1,2}$ and $w_{y1,2}$ correspond to their respective arm widths. The variable $Z_L$ represents the reactive load connected to the parasitic crossed-dipole. The defined bounds ensure physically realizable dimensions and maintain the a compact overall structure.

The MOGA in HFSS was selected for its capability to efficiently explore large, design spaces and avoid premature convergence to local optima. A population of 40 candidate geometries was evolved across 60 generations, with each candidate's fitness determined by the computed $G_{\text{LHCP}}$ extracted directly from full-wave simulations. The final optimized geometrical parameters obtained from this process are summarized in Table~\ref{tab:optimized_params}.

\begin{table}[!h]
\centering
\caption{Optimized Parameters of the Crossed-Dipole Array @ 3.5 GHz}
\label{tab:optimized_params}
\vspace{-0.3cm}
\footnotesize{All dimensions are in millimetres (mm).}
\vspace{4pt}

\renewcommand{\arraystretch}{1.2}
\setlength{\tabcolsep}{6pt}
\begin{tabular}{cccccccc}
\hline
$l_{x1}$ & $l_{x2}$ & $l_{y1}$ & $l_{y2}$ & $w_{x1}$ & $w_{x2}$ & $w_{y1}$ & $w_{y2}$\\
\hline
35.21 & 43.83 & 40.17 & 44.64 & 2.95 & 4.14 & 3.30 & 3.29 \\
\hline
\end{tabular}
\end{table}

The optimal reactive load was found to be capacitive, with an impedance of $Z_L = -j74.96~\Omega$, corresponding to an equivalent capacitance of 
$C = 1 / (2\pi f |X_L|) = 0.61~\text{pF}$. This capacitive loading adjusts the phase of the induced current to achieve constructive interference in the $+z$ direction, enhancing the end-fire radiation.

\begin{figure*}[t]
    \centering
    \includegraphics[width=\textwidth]{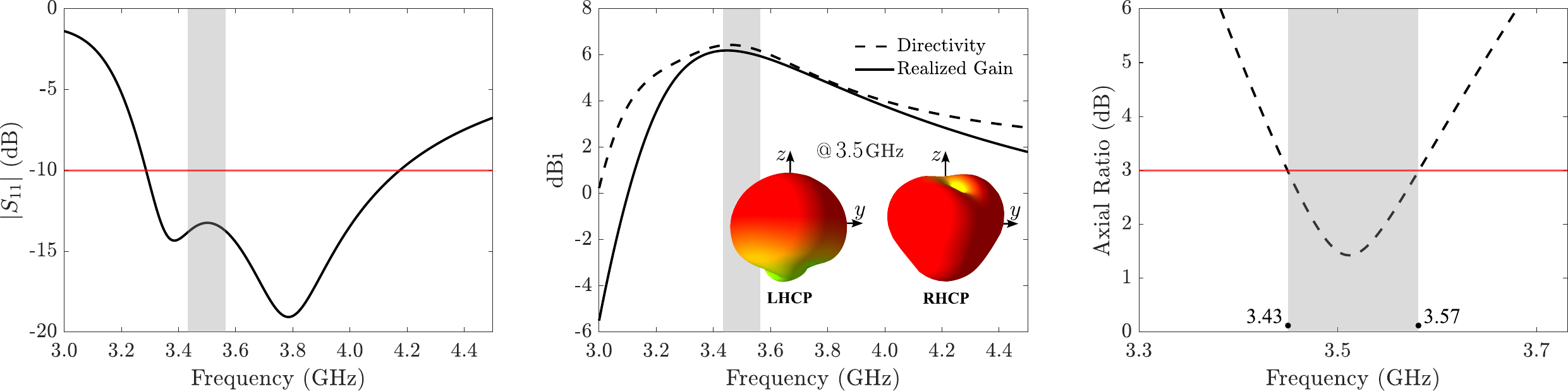}
    \caption{Simulated results of the optimized crossed-dipole array: 
  (a) reflection coefficient $|S_{11}|$, 
  (b) directivity and realized-gain, and 
  (c) axial ratio.}
    \label{fig:combined}
\end{figure*}

\section{Simulation Results and Analysis}

The proposed antenna demonstrates impedance matching and polarization purity, as evidenced in Fig. \ref{fig:combined}. Specifically, the simulated reflection coefficient remains below --10\,dB from 3.29 to 4.17\,GHz, corresponding to an impedance bandwidth of 23.75\%. At the target frequency of 3.5\,GHz, the antenna maintains good matching with $|S_{11}|\approx-13.2$\,dB and an input impedance close to $50~\Omega$. 
The axial-ratio (AR) response confirms stable LHCP, remaining below the 3\,dB threshold between 3.43 and 3.57\,GHz, which corresponds to an AR bandwidth of 4\%. This is also the \textit{common operating band}, where both impedance matching and polarization purity are simultaneously achieved. 
Within this common operating band, the antenna provides an LHCP realized-gain of approximately 6.14\,dB at 3.5\,GHz in the end-fire $+z$ direction, while the RHCP component is suppressed below --4.3\,dB, confirming efficient and polarization-pure radiation along the intended axis.

To benchmark the achieved realized gain relative to the physical aperture, the electrical size of the array was evaluated from the minimum enclosing sphere encompassing both crossed-dipole elements.
This yields an effective radius of $a \approx 22.6$~mm at 3.5~GHz, corresponding to an electrical size of $ka \approx 1.65$. The overall end-fire dimension of the array, measured along the $z$-axis, is approximately 13.7~mm, highlighting the low-profile nature of the structure in the radiation direction. According to Harrington's directivity limit~\cite{Harrington1958}, given by $D_{\max} = (ka)^2 + 2ka$, the theoretical maximum directivity for this electrical size is approximately 7.9~dBi. The simulated peak realized gain of 6.14~dB is therefore within about 1.8~dB of this bound, confirming that the proposed array approaches the maximum achievable directivity for its size while preserving good impedance matching and circular polarization purity.

\section{Conclusion}

A superdirective two-element single-feed crossed-dipole array capable of producing left-hand circularly polarized end-fire radiation at 3.5~GHz has been presented. Through careful adjustment of the strip dipole dimensions (lengths and widths) and the incorporation of a reactive load connected to the parasitic element, the antenna attains superdirective performance and circular polarization within a compact electrical size, while preserving excellent impedance matching and polarization purity.

The optimized configuration exhibits a common impedance and axial-ratio bandwidth of 3.49--3.57~GHz, and a realized gain of 6.1~dB, for $ka\approx1.65$. These results demonstrate that superdirectivity and circular polarization can be achieved concurrently in a simple, low-profile array suitable for integration into compact sub-6~GHz wireless platforms.

\section*{Acknowledgment}
This work was supported by the Program ``MEDICUS'' of the University of Patras, Greece.

\balance


\begin{thebibliography}{16}

\bibitem{Hansen1981Fundamental}
R.~C. Hansen, ``Fundamental limitations in antennas,'' \emph{Proceedings of the IEEE}, vol.~69, no.~2, pp.~170--182, 1981.

\bibitem{uzkov}
A.~I. Uzkov, ``An approach to the problem of optimum directive antenna design,'' \emph{Comptes Rendus (Doklady) de l'Academie des Sciences de l'URSS}, vol.~53, pp.~35--38, 1946.

\bibitem{Assimonis2023How}
S.~D. Assimonis, ``How challenging is it to design a practical superdirective antenna array?'' in \emph{2023 IEEE SENSORS}, Vienna, Austria, 2023, pp.~1--4.

\bibitem{Altshuler2005monopole}
E.~E. Altshuler, T.~H. O'Donnell, A.~D. Yaghjian, and S.~R. Best, ``A monopole superdirective array,'' \emph{IEEE Transactions on Antennas and Propagation}, vol.~53, no.~8, pp.~2653--2661, 2005.

\bibitem{Yaghjian2008esefa}
A.~D. Yaghjian, T.~H. O'Donnell, E.~E. Altshuler, and S.~R. Best, ``Electrically small supergain end-fire arrays,'' \emph{Radio Science}, vol.~43, no.~3, 2008.

\bibitem{jiao2018antenna}
R.~Jiao, Y.~Sun, J.~Sun, Y.~Jiang, and S.~Zeng, ``Antenna design using dynamic multi-objective evolutionary algorithm,'' \emph{IET Microwaves, Antennas \& Propagation}, vol.~12, no.~13, pp.~2065--2072, 2018.

\bibitem{Goudos2013moa}
S.~K. Goudos, K.~A. Gotsis, K.~Siakavara, E.~E. Vafiadis, and J.~N. Sahalos, ``A multi-objective approach to subarrayed linear antenna arrays design based on memetic differential evolution,'' \emph{IEEE Transactions on Antennas and Propagation}, vol.~61, no.~6, pp.~3042--3052, 2013.

\bibitem{ibrahim2023design}
S.~K. Ibrahim, M.~J. Singh, S.~S. Al-Bawri, H.~H. Ibrahim, M.~T. Islam, M.~S. Islam, A.~Alzamil, and W.~M. Abdulkawi, ``Design, challenges and developments for 5G massive MIMO antenna systems at sub 6-GHz band: A review,'' \emph{Nanomaterials}, vol.~13, no.~3, p.~520, 2023.

\bibitem{chen20235g}
W.~Chen, X.~Lin, J.~Lee, A.~Toskala, S.~Sun, C.~F. Chiasserini, and L.~Liu, ``5G-advanced toward 6G: Past, present, and future,'' \emph{IEEE Journal on Selected Areas in Communications}, vol.~41, no.~6, pp.~1592--1619, 2023.

\bibitem{ta2015crossed}
S.~X. Ta, I.~Park, and R.~W. Ziolkowski, ``Crossed dipole antennas: A review,'' \emph{IEEE Antennas and Propagation Magazine}, vol.~57, no.~5, pp.~107--122, 2015.

\bibitem{yepes2019analysis}
C.~Yepes, E.~Gandini, S.~Monni, F.~E. van Vliet, A.~Neto, and D.~Cavallo, ``Analysis and design of arrays with tilted directive dipole elements,'' in \emph{2019 49th European Microwave Conference (EuMC)}, 2019, pp.~17--20.

\bibitem{toh2003}
B.~Y. Toh, R.~Cahill, and V.~F. Fusco, ``Understanding and measuring circular polarization,'' \emph{IEEE Transactions on Education}, vol.~46, no.~3, pp.~313--318, 2003.

\bibitem{Lynch2024}
D.~P. Lynch, M.~M. Tentzeris, V.~Fusco, and S.~D. Asimonis, ``Super realized gain antenna array,'' \emph{IEEE Transactions on Antennas and Propagation}, vol.~72, no.~9, pp.~7030--7040, 2024.

\bibitem{lynch2024compactmeandereddipolearray}
D.~P. Lynch, M.~Zhao, A.~M. Graham, and S.~D. Asimonis, ``Compact meandered dipole array presenting super-realized gain,'' in \emph{2025 19th European Conference on Antennas and Propagation (EuCAP)}, 2025, pp.~1--4.

\bibitem{ansys_hfss}
{ANSYS, Inc.}, ``{ANSYS HFSS: High Frequency Structure Simulator},'' 2025, version 25.1, Canonsburg, PA, USA. [Online]. Available: \url{https://www.ansys.com/products/electronics/ansys-hfss}

\bibitem{Harrington1958}
R.~F. Harrington, ``On the gain and beamwidth of directional antennas,'' \emph{IRE Transactions on Antennas and Propagation}, vol.~6, no.~3, pp.~219--225, 1958.

\end{thebibliography}
\end{document}